\documentclass[aps,pra,amsmath,amssymb,reprint,superscriptaddress]{revtex4}

\usepackage{graphicx}% Include figure files
\usepackage{dcolumn}% Align table columns on decimal point
\usepackage{bm}% bold math
\begin{document}

\preprint{APS/123-QED}

\title{A transportable strontium optical lattice clock}% Force line breaks with \\

\author{N. Poli}
\author{M. Schioppo}
\altaffiliation[Present address: ]{National Institute of Standards and Technology, Time and Frequency Division, MS 847, Boulder, CO 80305, USA}%Lines break automatically or can be forced with \\
%\author{---}

\affiliation{Dipartimento di Fisica e Astronomia and LENS,
Universit\`a di Firenze and INFN Sezione di Firenze,
Via Sansone 1, 50019 Sesto Fiorentino, Italy}

\author{S. Vogt}
\author{St. Falke}
\altaffiliation[Present address: ]{TOPTICA Photonics AG, Lochhamer
Schlag 19, 82166 Graefelfing (Munich), Germany}
\author{U. Sterr}
\author{Ch. Lisdat}
\affiliation{Physikalisch-Technische Bundesanstalt, Bundesallee
100, 38116 Braunschweig, Germany}

\author{G. M. Tino}

\affiliation{Dipartimento di Fisica e Astronomia and LENS,
Universit\`a di Firenze and INFN Sezione di Firenze,
Via Sansone 1, 50019 Sesto Fiorentino, Italy}

\date{\today}% It is always \today, today,

\begin{abstract}
We report on a transportable optical clock, based on laser-cooled
strontium atoms trapped in an optical lattice. The experimental
apparatus is composed of a compact source of ultra-cold strontium
atoms including a compact cooling laser set-up and a transportable
ultra-stable laser for interrogating the optical clock transition.
The whole setup (excluding electronics) fits within a volume of
less than 2 m$^3$. The high degree of operation reliability of
both systems allowed the spectroscopy of the clock transition to
be performed with 10 Hz resolution. We estimate an uncertainty of
the clock of $7\times10^{-15}$.
%\begin{description}
%\end{description}
\end{abstract}

%\begin{abstract}
%An article usually includes an abstract, a concise summary of the work
%covered at length in the main body of the article.
%\begin{description}
%\end{description}
%\end{abstract}


%06.30.Ft Time and frequency
%06.20.fb Frequency standards
%37.10.Jk Optical cooling and trapping of atoms

\maketitle

%\tableofcontents

\section{\label{sec:1} Introduction}

%Advances in laser technology and manipulation of atoms have led to
%the realization of new atomic frequency standards based on optical
%transition with better stability and accuracy than their
%microwave-based predecessors.
As the development of optical clocks is currently pursued in many
laboratories worldwide \cite{Poli2013,Bloom2014,Chou2010,
Hinkley2013}, a broad range of applications is taking shape with
optical clocks on ground and in space employed for high precision
tests of fundamental physics \cite{Schiller2009, Wolf2009},
chronometric levelling-based geodesy \cite{Chou2010a}, improved RF
standards for navigation \cite{Fortier2011} and observations of
cosmic radio sources with very-long baseline interferometry (VLBI)
\cite{Rogers1983}. For all these purposes today's complex and
bulky optical-clock experimental setups need to be re-engineered
into more compact and power efficient systems, ensuring at the
same time a high stability and accuracy, but also high operation
reliability in critical environments \cite{Leibrandt2011}, such as
application in the field or even satellites.

A first step in the engineering challenge leading to space based
optical clocks is to demonstrate transportable clocks. These are
interesting because frequency comparisons of today's best optical
clocks cannot be done through satellite links and tests with
optical fiber links require dedicated equipment
\cite{Predehl2012,Williams2008,Calonico2014}. A frequency transfer
standard will allow for comparing clocks at the accuracy level
provided by the transportable clock \cite{Bize2005}.

In this paper we present the realization of a transportable
strontium optical clock, based on the integration of two
subsystems, a compact atomic source including a compact cooling
and trapping laser set-up, and a transportable clock laser,
providing the laser-cooled sample of strontium atoms and the
ultra-stable radiation source for interrogating the clock
transition, respectively. The clock laser, transported by van from
PTB in Braunschweig to LENS in Florence was employed to perform
high resolution spectroscopy of the clock transition
$^1$S$_0\,$-${}^3$P$_0$ of $^{88}$Sr in Lamb-Dicke regime and to
characterize systematic frequency shifts of the optical
transition.

In the following are presented in detail novel design solutions
employed for each subsystem that allowed us to reduce size, weight
and power consumption with respect to a traditional laser cooling
apparatus.

%The sample of cold $^{88}$Sr atoms loaded into a lattice at the
%magic wavelength was provided by a compact laser-cooling strontium
%source developed in Florence for transportable applications.

\section{\label{sec:2} The transportable clock setup}

\subsection{\label{subsec:strontium_source} Compact system for cooling and trapping strontium atoms}
The compact laser-cooling strontium source %(see Fig.\ref{fig:Strontium_source_assembly})
mainly consists of the following modules: the cooling laser
sources, a frequency distribution breadboard and a vacuum system.
The three modules are connected through optical fibers and mounted
directly on a 120$\,$cm$\,\times\,$90$\,$cm breadboard. The
control electronics is hosted under the main breadboard in a 19"
rack (size 60$\,$cm$\,\times\,$60$\,$cm$\,\times\,$180$\,$cm).

%\begin{figure}[b]
%\includegraphics{Strontium_source_assembly.eps}
%\caption{\label{fig:Strontium_source_assembly} Picture of the
%compact laser-cooling strontium source with the details of the
%lasers set, optical distribution breadboard, vacuum system and
%electronics.}
%\end{figure}

%\subsubsection{Laser sources}

\begin{figure}[t]\begin{center}
\includegraphics[width=0.5\textwidth]{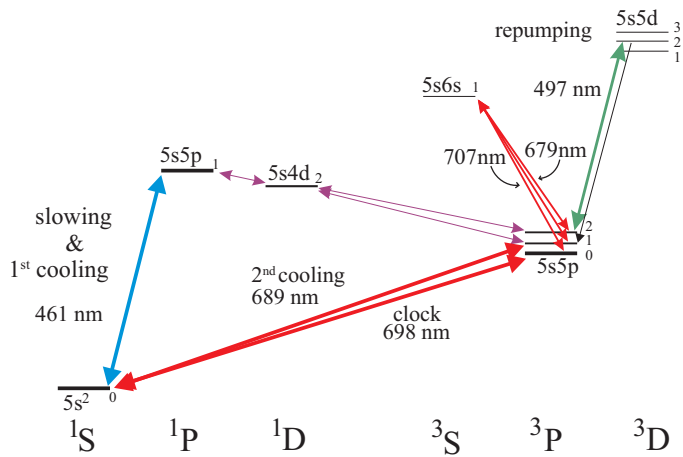}
\caption{Energy levels and relevant optical transition for neutral
Sr. The number near the level is the total angular momentum
$J$.\label{fig.LivelliSr}}
\end{center}
\end{figure}

Cooling and trapping of strontium atoms is performed through a
two-stage magneto-optical-trap (MOT) on the dipole allowed
$^1$S$_0\,$-${}^1$P$_1$ transition and on the intercombination
line $^1$S$_0\,$-${}^3$P$_1$, respectively (see Fig.
\ref{fig.LivelliSr}). All the employed lasers are based on
semiconductor diodes \cite{Poli2007, Poli2009}. The laser set is
composed of a frequency doubled 300$\,$mW 461$\,$nm laser, a
50$\,$mW 689 nm laser, a 420$\,$mW 813$\,$nm (master + tapered
amplifier) laser and two $\sim$20$\,$mW 679$\,$nm and 707$\,$nm
repumper lasers. These sources are used for the two-stage
laser-cooling and subsequent optical trapping in a one-dimensional
(1D) lattice at the magic wavelength \cite{Katori2009}.

%The laser source at 689$\,$nm for the second laser-cooling stage
%on the $^1$S$_0\,$-${}^3$P$_1$ transition is a laboratory system
%based on diode lasers, pre-stabilized on a $\sim10000$ finesse
%cavity \cite{Ferrari2003, Sorrentino2006, Poli2006} and frequency
%referenced to the atomic transition through saturation
%spectroscopy. This laser is located in a different laboratory and
%is coupled to the compact laser-cooling system through a 200$\,$m
%long optical fiber, after which the needed optical power level is
%obtained by injecting a 50$\,$mW slave laser.

%\subsubsection{Frequency distribution breadboard}
All the optical beams for the first laser-cooling stage at 461 nm
are produced in a compact module (size
30$\,$cm$\,\times\,$40$\,$cm$\,\times\,$8$\,$cm). %see Fig.\ref{fig:Strontium_source_assembly}
The module contains ultra-stable mountings for mirrors,
polarization cubes, plates, lenses and acousto-optic-modulators
(AOMs) to which the 461 nm radiation is delivered by fibre
\cite{SchioppoPhDThesis}.

\begin{figure}[t]
\includegraphics[width=0.6 \textwidth]{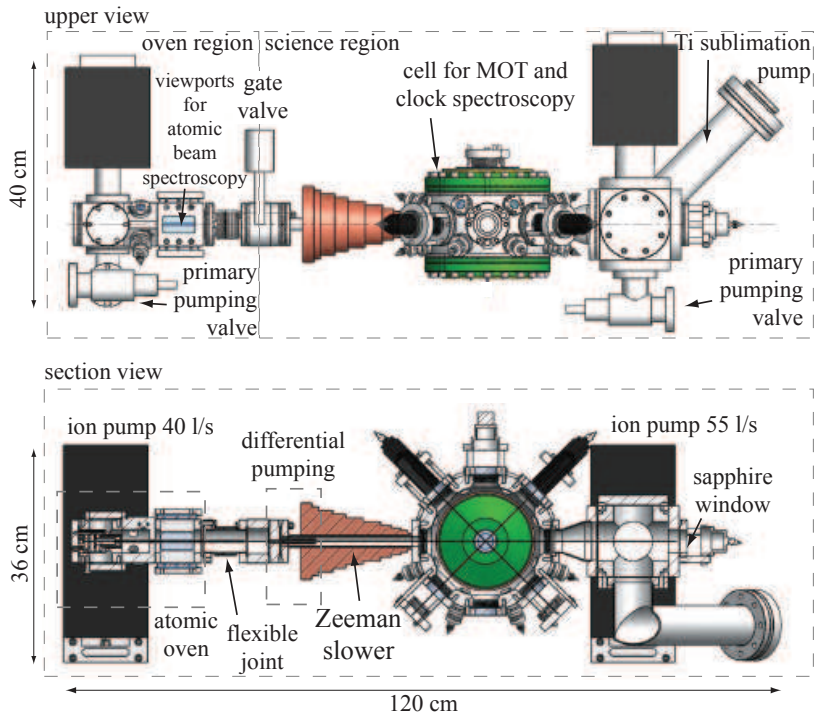}
\caption{\label{fig:vacuum_system}  Technical drawing of the
vacuum system (upper view and section side view) showing the main
part of the system (the oven region, on the left and the science
region, on the right.).}
\end{figure}

In order to reduce the number of beam shaping optical elements
used in the breadboard while maintaining a high diffraction
efficiency on AOMs, an input beam $1/e^2$ diameter has been set to
$0.5\,$mm. As a result we obtain high diffraction efficiency on
AOMs ($>$80 \% in single pass configuration) without the use of
any additional telescope, while still keeping the possibility of
resolving the diffraction orders at short distances. Moreover with
this choice a coupling efficiency into optical fibers of about 65
\% is achieved.  Then, the overall efficiency of a typical $\sim
25\,$cm long optical path inside the breadboard, including the AOM
diffraction efficiency (in single-pass configuration) and the
coupling efficiency into the output fiber, is typically
$\sim50\,\%$. The frequency detunings from the
$^1$S$_0\,$-${}^1$P$_1$ resonance for Zeeman slower and MOT beams,
of $-320\,$MHz and $-40\,$MHz respectively, are obtained by using
two AOMs in single-pass configuration, driven at 170 MHz ($-$
order) and 110 MHz ($+$ order), respectively. All output beams can
also be completely shut off by fast and compact mechanical
shutters.
%\begin{figure}[b]
%\includegraphics{distribution_breadboard_assembly.eps}
%\caption{\label{fig:distribution_breadboard} Picture of the
%compact fiber-coupled distribution breadboard for the first
%laser-cooling stage on the $^1$S$_0\,$-${}^1$P$_1$ transition at
%461$\,$nm (dimensions:
%30$\,$cm$\,\times\,$40$\,$cm$\,\times\,$8$\,$cm). The entire set
%of frequency required for Doppler cooling are produced inside this
%breadboard by the use of acousto-optical modulators (AOM) in
%single-pass and double-pass configuration.}
%\end{figure}
A beam for frequency stabilization of the first stage cooling
light is obtained from a double-pass configuration through an AOM
driven at 75$\,$MHz and sent to the atomic beam used to load the
MOT (see Fig. \ref{fig:vacuum_system}). In order to extract an
error signal from spectroscopy on the cooling transition, the
driving RF signal is also frequency modulated at $\sim$ 10$\,$kHz
with a peak to peak deviation of $\sim$10 MHz. A low power ($\sim$
1 mW) resonant beam for absorption imaging is generated in a
similar double pass configuration.  For both double-pass AOMs a
cat's-eye configuration has been realized by focusing the optical
beam on the retro-reflecting mirror with a lens at distance f = 60
mm from AOM and retro-reflection mirror.
%In conclusion, on Fig.
%\ref{fig:distribution_breadboard} are reported the typical power
%levels obtained for each beam coupled on the output fiber.
%values Then, starting from the 250$\,$mW
%optical power available at the breadboard input, six
%fiber-collimated output beams are generated, respectively for
%frequency stabilization of the laser on the cooling transition
%(1$\,$mW after fiber), absorption imaging (200$\,\mu$W), Zeeman
%slowing (30$\,$mW), magneto-optical trapping (three outputs:
%20$\,$mW, 15$\,$mW and 15$\,$mW).

%\subsubsection{Vacuum system}
The atomic sample is trapped in a vacuum system (size
1200$\,$cm$\,\times\,$40$\,$cm $\times$ 36$\,$cm),  see Fig.
\ref{fig:vacuum_system}) consisting of the oven region, where a
collimated Sr atomic beam is produced, a Zeeman slower and the
science region where the Sr atoms are subsequently Zeeman slowed,
trapped in a two-stage MOT and eventually transferred in a 1D
vertical optical lattice for clock spectroscopy. The oven region
is pumped by a 40 l/s ion pump (pressure during operation $\sim
10^{-6}$ Pa), the science region is evacuated by a 55 l/s ion pump
and a titanium sublimation pump (pressure $\sim10^{-7}$ Pa). A
differential pumping tube (internal diameter 5 mm, length 7.5 cm)
ensures to maintain the pressure difference between the two
regions.
%A gate
%valve is placed between the two regions so that they can be
%evacuated (or opened) through two dedicated primary pumping
%valves.

In the oven region, atoms are sublimated by a compact and
efficient dispenser \cite{Schioppo2012} based on a heater placed
in vacuum, providing an atomic flux intensity of
$1.7\times10^{13}$ s$^{-1}$sr$^{-1}$ ($^{88}$Sr atoms) at the oven
temperature of $380\,^{\circ}$C with a total power consumption of
26 W. The atomic beam is collimated by a nozzle of about 120
capillaries (internal diameter 200$\,\mu$m, length $8\,$mm). The
high collimation ($\sim 40$ mrad) and flux allow the laser
stabilization to be performed with high signal-to-noise employing
a simple transverse fluorescence spectroscopy on the atomic beam
in the oven chamber (see Fig. \ref{fig:vacuum_system}).
%Thus the
%complication of saturation spectroscopy in a dedicated vapor cell
%for stabilization of the 461$\,$nm laser is avoided.
The atomic beam propagates along a 23$\,$cm long tube externally
wrapped with coils for Zeeman slowing. The magnetic field shape
for Zeeman slowing has been designed to smoothly match the
off-axis component of the MOT coils' field (see Fig.
\ref{fig:slowing_dynamics_final}) \cite{SchioppoPhDThesis}. The
first part of the Zeeman slower (before the inversion of the field
given by the MOT coils) has a length of $18\,$cm and it operates
with a power consumption of 17$\,$W.

The decelerated atomic beam is trapped in a MOT at the center of a
science cell with eight CF40 and sixteen CF16 optical windows. Two
custom CF150 flanges host the pair of coils (in anti-Helmholtz
configuration as needed for MOT operation) outside the MOT chamber
2.6$\,$cm away from the atoms. In this configuration a magnetic
field gradient of $\sim 500$ mT/m is obtained with a total power
consumption of $\sim60$ W. With these levels of power consumption
for oven, Zeeman slower and MOT coils we could avoid the
complication of water cooling, with a typical chamber temperature
of 318 K.

%\begin{figure}[t]
%\includegraphics{slowing_dynamics_final.eps}
%\caption{\label{fig:slowing_dynamics_final}Simulation of the
%slowing process along the Zeeman slower. The field generated by
%the main Zeeman slower coils is matched with the off-axis field of
%the MOT field. The Zeeman slower beam detining is then adjusted in
%order to obtain a final velocity spectrum centered around 40 m/s,
%below the MOT capture velocity.}
%\end{figure}

To make the alignment of the MOT beams long-term stable, the
cooling beams at 461 nm and 689 nm are delivered by three dichroic
fiber-coupled beam expanders fastened to the science cell. %(see Fig.\ref{fig:dich_tel_assembly}
A system of setting screws and counter screws allows fine
alignment and locking of the MOT beams. Similarly, the MOT beam
retroreflectors, repumping, imaging and clock spectroscopy
output-couplers are fastened onto the science cell. The alignment
of the MOT beams has been maintained for more than one year
without any further adjustment.

%\begin{figure}[t]
%\includegraphics{dich_tel_assembly_2.eps}
%\caption{\label{fig:dich_tel_assembly} Photo (left) and optical
%scheme (right) of the three dichroic fiber-coupled beam expanders
%employed to deliver the optical radiation at 461$\,$nm and
%689$\,$nm respectively for the first and second laser-cooling
%stage MOT.}
%\end{figure}

\subsection{\label{subsec:clock_laser} Transportable clock laser}
The main clock laser breadboard is based on two diode lasers in a
master-slave setup. The master is a filter stabilized extended
cavity laser with resonator length 10 cm \cite{Baillard2006}.
About 0.5 mW of the master-laser power is used to injection lock
the slave laser. The breadboard has a total size of 60 cm $\times$
45 cm $\times$ 10 cm, its design is similar to frequency
distribution breadboard presented in the previous section. A 200
MHz AOM in double-pass configuration is used to bridge the
frequency gap between the optical resonator with free spectral
range of 1.5 GHz and the frequency of the clock transition. This
AOM is used to scan the frequency of the laser and also to enable
frequency feedback when the laser is locked to the clock
transition. Two output ports for clock spectroscopy and for
counting of the laser frequency provide a maximum optical power of
about 2 mW each. Both ports are provided with AOMs for switching
which can also be used to stabilize the optical path length of the
fibers. The interferometric setup for this stabilization is
included on the breadboard \cite{Falke2012}. The laser was locked
to a transportable high finesse cavity whose transportability had
already been demonstrated within the SOC project \cite{Vogt2011}.

\section{\label{sec:results} Experimental Results}

\subsection{Cooling and trapping}
\begin{figure}[t]
\includegraphics[width=0.6 \textwidth]{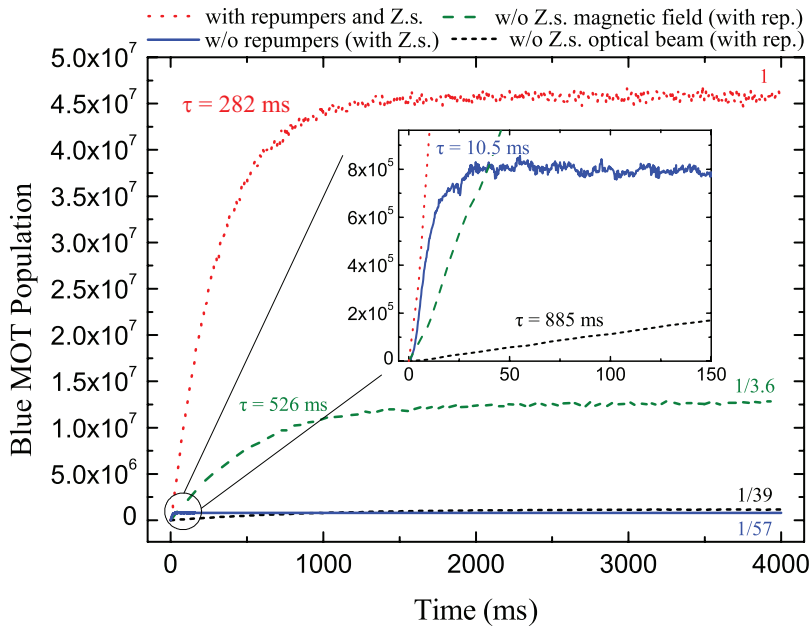}
\caption{\label{fig:BlueMOT_loading_cropped} Comparison among blue
MOT loading curves and final blue MOT population obtained with
repumpers and Zeeman slower beam (red dash), without Zeeman slower
magnetic field (green dash), without repumpers (blue line) and
without Zeeman slower beam (black dash).}
\end{figure}

In order to develop a more compact and low power consumption
experimental system for the production of ultra-cold strontium
atoms, several design solutions have been adopted, together with
an optimization of the efficiency of each cooling and trapping
step. In order to slow down the atoms sublimating from the oven
from 430$\,\text{m}/\text{s}$ to below 130$\,\text{m}/\text{s}$,
an optical beam at $461\,$nm, circularly polarized, is sent
counter propagating to the atomic beam. The laser beam has an
optical power of $30\,$mW, an initial $1/e^2$ radius of $r =
5\,$mm, focused after 1$\,$m to compensate the absorption of
photons in the slowing process. Atoms are kept in resonance during
the slowing by compensating the variation of Doppler shift with
the proper Zeeman shift from the magnetic field provided by the
Zeeman slower. The sample of cold $^{88}$Sr atoms is produced
through two laser-cooling stages. The first stage consists of a
MOT operating on the $^1$S$_0\,$-${}^1$P$_1$ transition at
$461\,$nm. This so called blue MOT is realized by three pairs of
counter-propagating optical beams, circularly polarized, detuned
by $\delta_L=-2\pi\times40\,$MHz, with a saturation parameter for
each beam of $s\sim1$ and a magnetic field gradient of $500\,$
mT/m. The blue MOT capture velocity is $
v_{\text{c}}=\sqrt{2a_{\text{max}}r}\simeq100\,\text{m/s}$ where
$a_{\text{max}}=\frac{1}{M}\hslash
k_{L}\frac{\Gamma}{2}\frac{s}{1+s}$ is the maximum acceleration
exerted by cooling light, with  $k_L=2\pi/\lambda$ the wavevector
of the photons at $\lambda=461\,$nm, $\Gamma=2\pi\times32\,$MHz
the natural linewidth of the $^1$S$_0\,$-${}^1$P$_1$ transition
and $M$ the atomic mass of strontium 88. As expected we find that
the Zeeman slower increases the loading rate of the MOT, resulting
in a net increase of the blue MOT population by a factor of
$\sim40$ (see Fig. \ref{fig:BlueMOT_loading_cropped}).
% The blue MOT velocity has
%to be compared with the most probable velocity in the atomic beam
%produced by the oven at $T=380\,^{\circ}$C, $
%v_{\text{beam}}=\sqrt{k_{\text{B}}T/M}\simeq430\,\text{m}/\text{s}$,
%with $k_\text{B}$ Boltzmann constant.

The $^1$S$_0\,$-${}^1$P$_1$ transition used for the blue MOT is
not perfectly closed due to the decay channel of the
5p$\,{}^1$P$_1$ state towards the 4d$\,{}^1$D$_2$ state, which has
a lifetime of $0.5$ ms \cite{Xu2003}. Atoms in the latter state
may decay to the ground state through the 5p$\,{}^3$P$_1$ within
less than 1$\,$ms or may decay to the metastable 5p$\,{}^3$P$_2$
state and be lost. In order to recycle the atoms stored in the
metastable 5p$\,^3$P$_2$ state a 10 mW repumper laser at 707$\,$nm
is used to pump these atoms in the 6s$\,^3$S$_1$ state. An
additional 10 mW laser at 679$\,$nm is necessary to deplete the
$^3$P$_0$ state since it is also coupled to the 6s$\,^3$S$_1$
state. The repumping laser beams are superimposed, expanded to 10
mm of diameter and sent to the blue MOT and retroreflected. The
repumping increases the atom number in the blue MOT by a factor
$\sim60$. We studied the loading dynamics of the blue MOT
population $N$, given by $dN/dt=\phi_{c}-\Gamma_{L}N\,-\beta'
N^2$, where $\phi_{c}$ is the effective loading rate of atoms in
the MOT, $\Gamma_{L}$ is the linear loss rate (mainly due to
background collisions when repumpers are operating) and $\beta'$
is the the coefficient for two-body collisional loss
\cite{Dinneen1999}. The previous equation with the initial
condition $N(t=0)=0$ has the standard solution $
N(t)=N_{st}(1-\exp(-t/\tau))/(1-\xi\exp(-t/\tau))\,$ where
$N_{st}$ is the stationary number of trapped atoms, $\tau$ is the
effective trap loading time and $\xi$ is the collisional loss
fraction \cite{Dinneen1999}. With repumpers we measured
$N_{st}=4.5\times10^7$ $^{88}$Sr atoms, corresponding to an atomic
density of $n_0=8\times10^9$ cm$^{-3}$, $\tau=282\,$ms and
$\beta'=1\times10^{-8}$ s$^{-1}$, leading to a rate of captured
atoms of
$\phi_{c}=N_{st}/\tau+\beta'N{_{st}}^2=9.7\times10^{7}\,\text{s}^{-1}$.
Considering that the rate of atoms effused by the oven in the
solid angle covered by the Zeeman slowing beam is
$6.5\times10^8\,\text{s}^{-1}$, about $15$ \% of the atoms are
actually trapped into the blue MOT. This efficiency is mostly
determined by the optical pumping into the 4d$\,{}^1$D$_2$ state,
where atoms can be considered lost from the slowing dynamics since
the state lifetime is comparable to the slowing time scale. The
probability of an atom of decaying into the ground state after the
absorption of $n$ photons is
\begin{figure}[t]
\includegraphics{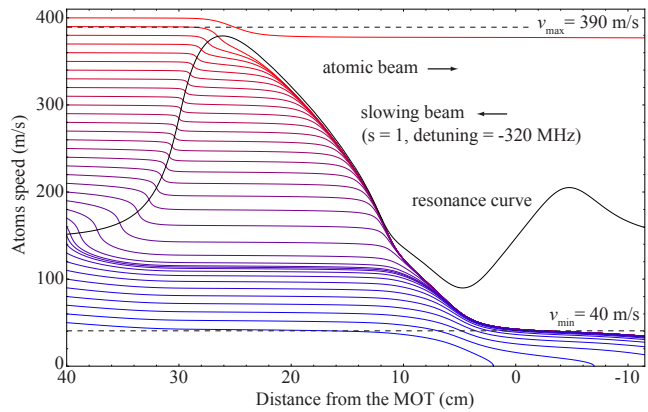}
\caption{\label{fig:slowing_dynamics_final} Simulation of the
slowing dynamics with the field profile given by the Zeeman slower
coils and the off-axis quadrupole field of the MOT coils.}
\end{figure}
\begin{equation}
\left(\frac{\Gamma_{\text{b}}}{\Gamma_{\text{b}}+\Gamma_{\text{d}}}\right)^{n}\simeq\left(1-\frac{\Gamma_{\text{d}}}{\Gamma_{\text{b}}}\right)^{n}\simeq\exp\left(-\frac{\Gamma_{\text{d}}}{\Gamma_{\text{b}}}n\right)\,.
\end{equation}
where $\Gamma_b=2.0\times10^8$ s$^{-1}$ and
$\Gamma_d=3.9\times10^3$ s$^{-1}$ are the decay rates of the state
$^1$P$_1$ into $^1$S$_0$ and $^1$D$_2$ states, respectively.
Therefore the actual fraction of slowed atoms can be estimated
through
\begin{equation}
\eta\equiv\int_{v_{\text{min}}}^{v_{\text{max}}}
\exp\left(-\frac{\Gamma_{\text{d}}}
{\Gamma_{\text{b}}}n(v)\right)f(v)dv\simeq20\,\%\;,
\label{eq:Zeeman_slower_efficiency}
\end{equation}
where $v_{\text{min}}=40\,\text{m/s}$ is the minimum final
velocity of atoms at the end of the slowing dynamics,
$v_{\text{max}}=390\,\text{m/s}$ is the maximum atom velocity that
can be slowed down (according to the results of numerical
simulation shown in Fig. \ref{fig:slowing_dynamics_final}),
$n(v)=(v-v_{\text{min}})/v_{\text{rec}}$ is the number of photons
necessary to reduce the atom velocity from $v$ to
$v_{\text{min}}$, $v_{\text{rec}}=h/\lambda
M\simeq1\,\text{cm}/\text{s}$ is the recoil velocity due to the
absorption of a photon at $461\,$nm and $f(v)$ is the velocity
distribution in an effusive atomic beam at $T=380\,^{\circ}$C
\begin{equation}
f(v)=2\left(\frac{M}{2k_{\textrm{B}}T}\right)^{2}v^{3}\exp\left(-\frac{Mv^{2}}{2k_{\textrm{B}}T}\right)\;.
\end{equation}

Therefore the estimation of the Zeeman slower efficiency
$\eta\simeq20\,\%$ given by Eq. \ref{eq:Zeeman_slower_efficiency}
is close to the measured value. Additionally the Blue MOT
population enhancement factor due to the operation of Zeeman
slower can be estimated from the ratio between the actual fraction
of slowed atoms $\eta$ and the fraction of trapped atoms without
Zeeman slower
\begin{equation}
\eta\;/\int_{0}^{v_{\text{c}}}f(v)dv\simeq70\;,
\end{equation}
which is close to the measured value of $\sim40$.

\begin{figure}[t]
\begin{center}
%\includegraphics[width=0.42\textwidth]{LaserCooling1.eps}
%\hspace{0.5cm}
\includegraphics[width=0.6\textwidth]{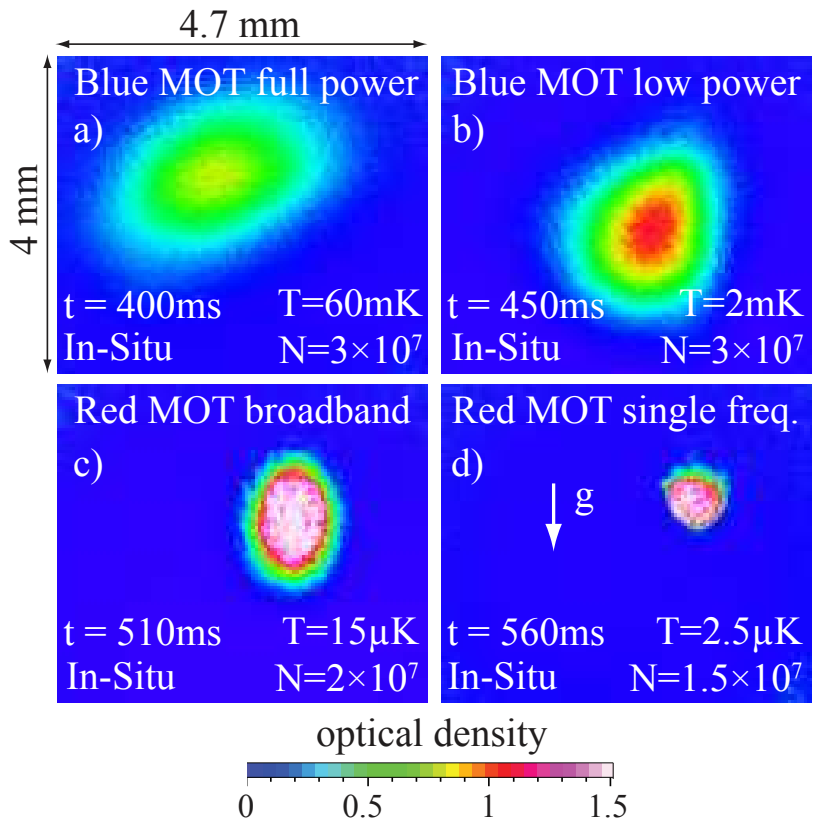}
\caption{\label{fig:laser_cooling} Absorption imaging
\emph{in-situ} of the atomic sample at the relevant phases of the
laser cooling sequence lasting a time $t$, with a constant blue
MOT loading time of 400 ms. a) blue MOT full power. b) blue MOT
low power. c) red MOT broadband. d) red MOT single frequency.}
\end{center}
\end{figure}

The final temperature of the blue MOT is minimized by continuously
reducing the optical power of the cooling beams from the initial
total saturation parameter of $s=1$ to a final value of
$s=5\times10^{-3}$ in $10\,$ms. This power-reduced phase lasts for
50$\,$ms, leading to a final temperature of the blue MOT of about
$\sim2\,$mK measured with absorption imaging (see Fig.
\ref{fig:laser_cooling}).

%\subsubsection{Second laser-cooling stage: red MOT}

The second laser-cooling stage is performed by using the
$^1$S$_0\,$-${}^3$P$_1$ intercombination transition at 689$\,$nm
(red MOT). The optical beams needed for this phase are
superimposed on the blue MOT beams through the three dichroic beam
expanders. The $1/e^2$ diameter of the beams for the red MOT is
10$\,$mm and the total intensity incident on the atoms is
$60\,\text{mW}/\text{cm}^2$. The atomic sample at the end of the
first cooling stage ($T\sim2\,$mK) has a Doppler width of
$\sim2\,$MHz, too large to be efficiently transferred into the red
MOT operating on a $7\,$kHz natural linewidth transition. For this
reason the spectrum of the 689$\,$nm laser is artificially
broadened \cite{Katori1999,Loftus2004}. This is realized by
modulating the radio frequency driving of the AOM setting the
frequency detuning of the cooling beams. The modulation frequency
is $50\,$kHz, with a span of $4\,$MHz, leading to an optical
spectrum of $\sim$ 80 sidebands, with a saturation parameter of
400 for each one, with the closest to resonance by $-300\,$kHz.
This so called red MOT broadband cooling phase lasts for 60$\,$ms
and more than 70$\,\%$ of the blue MOT population is transferred
into the red MOT. At the beginning of this phase the Zeeman slower
field is turned off and the MOT magnetic field gradient is only 20
mT/m in order to have all the atomic velocity classes resonant
with the cooling light and it is linearly increased to 70 mT/m in
a time interval of $30\,$ms to compress the atomic sample. With a
blue MOT loading time of $400$ ms the final population of the
broadband red MOT is $2\times10^7$ atoms, with an atomic density
of $1.5\times10^{11}\,\text{cm}^{-3}$  and a temperature of
$\sim15\,\mu$K.

The second laser-cooling phase is completed by employing a single
frequency red MOT, with a detuning of $-300\,$kHz, a constant
magnetic field gradient of 20 mT/m and reducing the total
intensity of the cooling beams down to
$500\,\mu\text{W}/\text{cm}^2$ in $50\,$ms. The single frequency
red MOT phase produces a sample of $1.5\times10^7$ atoms, with an
atomic density of $2\times10^{11}\,\text{cm}^{-3}$ and a
temperature of $\sim2.5\,\mu$K. Since the gravity force is
comparable to the radiation force of the red MOT beams, the atomic
sample sags down from the center of the MOT quadrupole field,
assuming an half-disk-like shape with a vertical and horizontal
diameter of $500\,\mu$m \cite{Loftus2004}.

%\subsection{Vertical lattice trapping}

In order to have a long Doppler-free interrogation time of the
clock transition, the laser-cooled strontium sample is trapped
into a vertical lattice realized by retroreflecting a $290\,$mW
laser beam near the magic wavelength for $^{88}$Sr
$\lambda_\text{magic}=813.42757(62)\,$nm \cite{Akatsuka2010}. At
this wavelength the light shifts of the $^1$S$_0$ and $^3$P$_0$
level are equal, so that the frequency of the clock transition
$^1$S$_0\,$-${}^3$P$_0$ is not light-shifted by the lattice. The
lattice laser output is coupled into a single mode fiber
delivering an optical beam with $1/e^2$ diameter of $1.2\,$mm,
which is expanded and focused into the red MOT by a $300\,$mm
focal length lens.  The latter is mounted on a 3-axis translation
stage with micro-metric actuators, so that the alignment of the
beam onto the red MOT can be finely tuned. The resulting beam
waist radius on the atomic cloud is $w_0\sim40\,\mu$m. After the
focus, the divergent lattice beam is then collimated and
retroreflected by means of a dichroic mirror. The latter is
employed to couple into the lattice the clock probe beam at
$\lambda_c=698\,$nm. The resulting beam diameter of the clock
light on the atoms is $74\,\mu$m. Taking into account the power
losses due to the telescope, focusing optics and cell windows the
estimated lattice trap depth is $U_0\simeq76\,E_{\text{rec}}$
(corresponding to $\sim12\,\mu$K, in temperature units), where
$E_{\text{rec}}=\hslash^2 k^2/2m$ is the photon recoil energy with
the wavevector $k=2\pi/\lambda_\text{magic}$ of the lattice light
and $m$ the atomic mass of $^{88}$Sr. At this depth of the
potential the estimated longitudinal trap frequency is
$\nu_z=2E_{\text{rec}}\sqrt{U_{0}/E_{\text{rec}}}/h=60\,$kHz.

The lattice is continuously operating and at the end of the second
laser-cooling stage about $3\times10^5$ atoms remain trapped,
populating $\sim1000$ sites, for a total extension of
$\sim400\,\mu$m corresponding to the vertical size of the red MOT.
The $1/e^2$ radius of the atomic cloud trapped in the lattice is
measured with absorption imaging to be $\sigma_r=11\,\mu$m. The
longitudinal $1/e^2$ radius of a single site, of the order of
$\lambda_L/2$, cannot be resolved by absorption imaging and is
estimated from the size
$\sigma_z=\sqrt{\hslash/m\omega_z}\sim45\,\text{nm}$ of the wave
function of atoms populating the ground longitudinal vibrational
state (is a good approximation since we have
$k_{\text{B}}T/h\nu_{z}\sim0.9$). Thus the trapping volume for
lattice site is
$V_{\text{site}}=(2\pi)^{3/2}\sigma_{r}^{2}\sigma_{z}\simeq5\times10^{-10}\,\text{cm}^{3}$.
In order to keep the shift and broadening effects on the clock
transition due to atomic collisions \cite{Lisdat2008}, we reduced
the number of trapped atoms to below $10^4$ (less than 10 atoms
per site) by reducing the blue total MOT loading time to 100 ms,
leading to a peak atomic density per site below
$2\times10^{-10}\,\text{cm}^{-3}$. The number of atoms is
controlled by varying the duration of the blue MOT stage and by
finely tuning the position of the gate valve separating the oven
from the science region.

The lifetime of atoms trapped into the lattice is measured to be
$1.4\,$s, limited by background gas collisions. Both power and
frequency of the lattice laser are not stabilized. The relative
RMS power fluctuation is below the $1$ \% level. The lattice
wavelength is tuned near to the magic wavelength of $^{88}$Sr
\cite{Akatsuka2010}, monitored by a wave-meter and measured to be
stable at $\lambda=813.4280(1)\,$nm over a time scale of several
hours.

\subsection{Lattice clock spectroscopy}
\begin{figure}[t]
\includegraphics[width=0.6\textwidth]{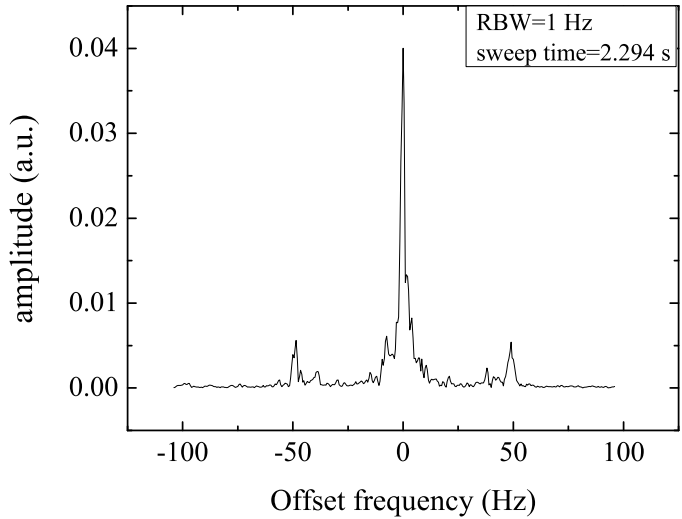}
\caption{\label{fig:beatclocklaser} Beatnote of the transportable
laser with a stationary clock laser at 698 nm. The estimated
emission linewidth for each laser is 1 Hz.}
\end{figure}

As a first test of the performance of the transportable clock
laser after the two-days, 1300 km-long transportation, the laser
was compared with a stationary clock laser \cite{Tarallo2011}.
Fig. \ref{fig:beatclocklaser} shows the beat note recorded just
after re-installation of the clock laser in the lab. The
installation process took about one day, mainly do to
re-thermalization of the transportable cavity, which was not
actively temperature stabilized during transportation. The beat
note shows a linewidth of the order of 1 Hz compatible with the
laser frequency stability of $2-3\times10^{-15}$ at 1 s.
The comparison with the stationary clock has also been used to
estimate the absolute frequency of the transportable clock laser
with an uncertainty of less than $1\,$MHz.
%Applying the proper
%correction extrapolated from the beating note measurement of the
%two clock lasers and from the detail of the frequency chain
%employed, the transportable clock laser was found to be resonant
%with the atomic transition already from the first scan.

\begin{figure}
\begin{center}
\includegraphics[width=0.9\textwidth]{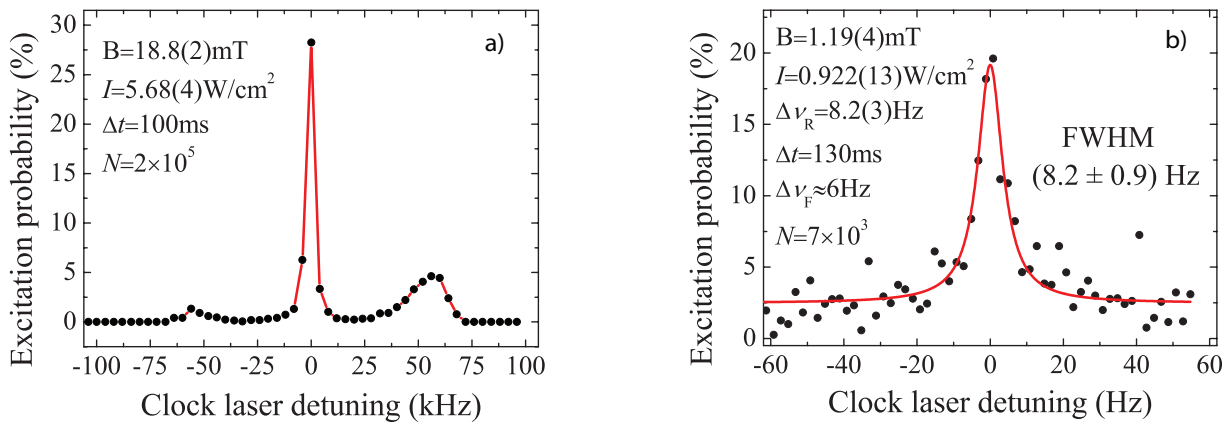}
\caption{\label{fig:clock_spectra} Spectra of the $^{88}$Sr clock
transition for different values of the mixing magnetic field
$\left|\mathbf{B}\right|$, clock probe beam intensity $I$ and
excitation pulse length $\Delta t$. The frequency axes have
arbitrary offsets. The clock resonance is fitted with a Lorentzian
function and the obtained FWHM is compared to the Fourier limit
linewidth $\Delta\nu_{\text{F}}\simeq0.8/\Delta t$ and to the Rabi
linewidth
$\Delta\nu_{\text{R}}=0.35\sqrt{\left|\Delta_{B}\Delta_{L}\right|}$,
where $\Delta_{B}$ and $\Delta_{L}$ are the second order Zeeman
shift and clock probe light shift, respectively (see the text). a)
shows a typical search-scan spectrum with the maximum number of
atoms loaded into the lattice $N\simeq2\times10^5$. b) is taken
with clock interrogation $\pi$ pulses and a lattice population of
$N=7\times10^{3}$.}
\end{center}
\end{figure}

\begin{figure}
\begin{center}
\includegraphics[width=0.4\textwidth]{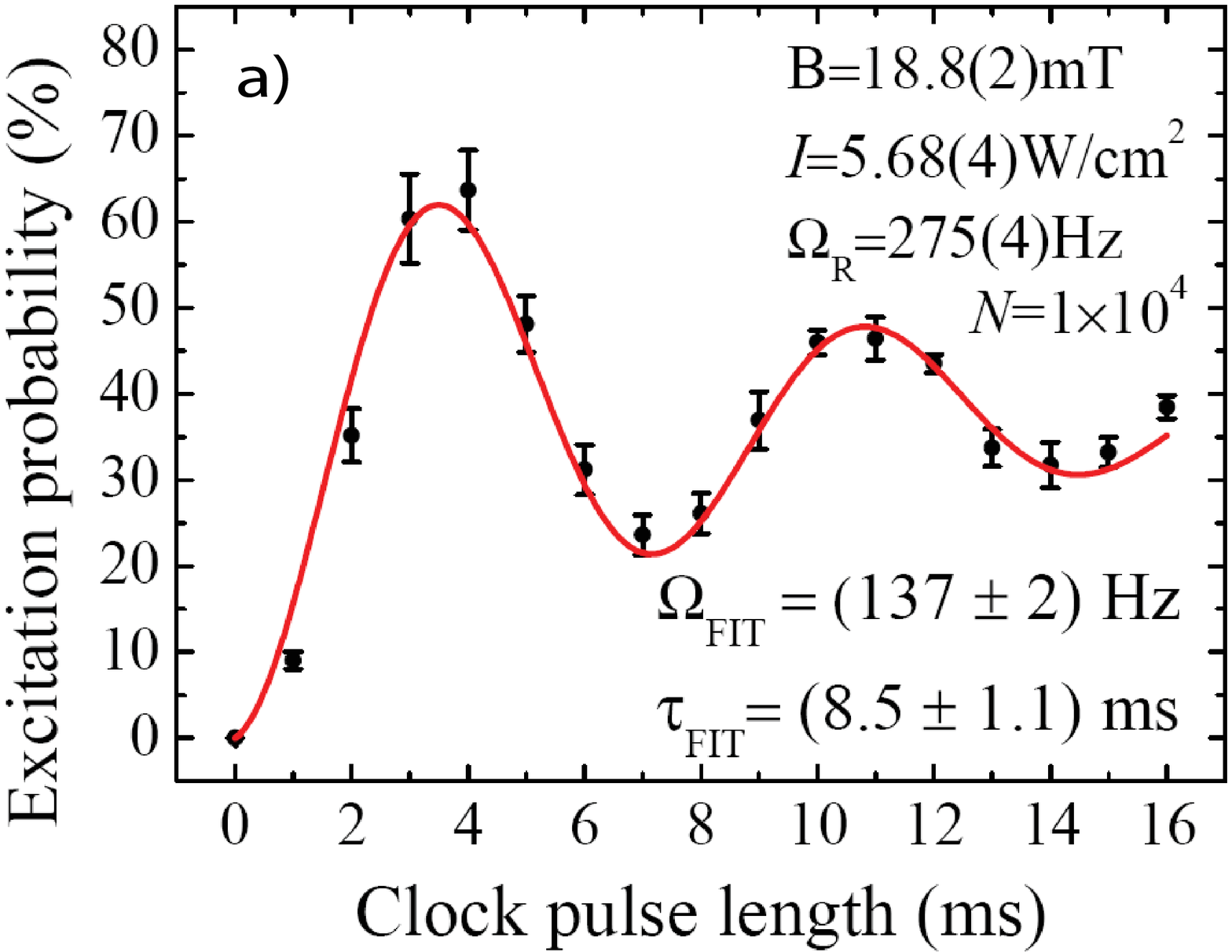}
\includegraphics[width=0.4\textwidth]{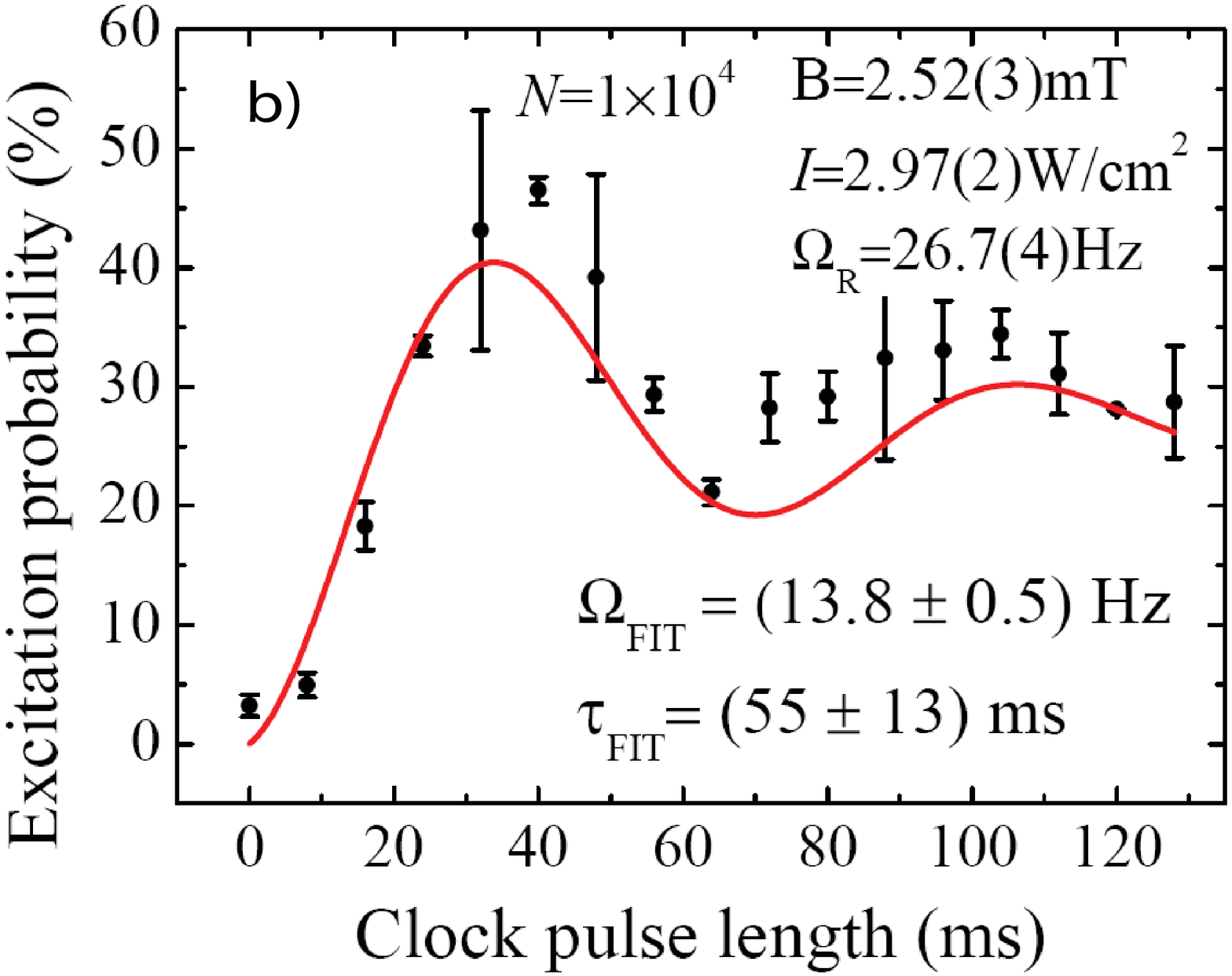}
\caption{\label{fig:Rabi_oscillations} Measurement of the Rabi
oscillations in condition of resonance with the clock transition
for different values of the mixing magnetic field and clock probe
beam intensity. The data are fitted with the function
$a(1-\cos(2\pi\Omega_{\text{FIT}}\Delta t)\exp(-\Delta
t/\tau_{\text{FIT}}))$, where $\Omega_{\text{FIT}}$ is the actual
atomic Rabi frequency and $\tau_{\text{FIT}}$ gives the
decoherence time-scale. The measured atomic Rabi frequency
$\Omega_{\text{FIT}}$ is $\sim50\,\%$ of the estimated Rabi
frequency $\Omega_{\text{R}}$ calculated from the mixing magnetic
field and clock laser intensity (see Eq. \ref{eq:Rabi}). The
corresponding $\pi$ pulse interrogation length $\Delta
t_{\pi}=1/2\Omega_{\text{FIT}}$ was employed for the clock spectra
b) of Fig. \ref{fig:clock_spectra}.}
\end{center}
\end{figure}

Considering that the single photon $^{88}$Sr clock transition
$^1$S$_0\,$-${}^3$P$_0$ is forbidden at any order, the
magnetic-field-induced spectroscopy method is used to controllably
allow the clock transition, by means of an external magnetic field
coupling the $^3$P$_0$ to the $^3$P$_1$ state
\cite{Taichenachev2006}. The search-scan is performed by using a
mixing magnetic field of $\left|\mathbf{B}\right|=19\,$mT and a
clock probe beam intensity of $I=5.7\,\text{W}/\text{cm}^2$,
leading to a Rabi frequency of $275\,$Hz, given by
\begin{equation}
\Omega_{\text{R}}=\alpha\sqrt{I}\left|\mathbf{B}\right|
\label{eq:Rabi}
\end{equation}
where $\alpha
(\text{Sr})=198\,\text{Hz}/(\text{T}\sqrt{\text{mW}/\text{cm}^{2}})$
\cite{Taichenachev2006}. In order to have a high-contrast spectrum
the excitation pulse length is $\Delta t=100\,$ms, thus
overdriving the clock transition having an estimated $\pi$ pulse
duration of $\Delta t_{\pi}=1.8\,$ms. In the search-mode the clock
laser frequency is changed by $4\,$kHz, covering a span of
$200\,$kHz in about 50$\,$s (experiment cycle 1$\,$s). A typical
search-mode scan is shown in Fig. \ref{fig:clock_spectra}. The
excitation probability is given by the ratio
$n(^{3}\text{P}_{0})/[n(^{1}\text{S}_{0})+n(^{3}\text{P}_{0})]$,
where $n(^{3}\text{P}_{0})$ and $n(^{1}\text{S}_{0})$ is the
atomic population of the $^{3}\text{P}_{0}$ and $^{1}\text{S}_{0}$
state, respectively. The $^{1}\text{S}_{0}$ population is obtained
through absorption imaging of the atomic sample after the clock
transition interrogation. Atoms in the $^{1}\text{S}_{0}$ state
are blown away from the lattice by the resonant 461$\,$nm imaging
beam. The atoms excited into the $^{3}\text{P}_{0}$ level by the
clock probe beam are pumped back into the $^{1}\text{S}_{0}$ state
by means of a 50$\,$ms pulse at $679\,$nm and $707\,$nm. The
$^{1}\text{S}_{0}$ population is then measured through an
additional absorption imaging sequence. From the spectra in Fig.
\ref{fig:clock_spectra}a, the motional sidebands are measured at
$\sim65$ kHz from the carrier
%, (compared with our
%initially estimated longitudinal lattice trapping frequency
%$\nu_z\simeq55\,$kHz)
, thus corresponding to a Lamb-Dicke parameter
$\eta=\sqrt{\nu_{R}/\nu_{z}}\simeq0.3$, where
$\nu_R=h/(2m\lambda^2)$ is the atomic recoil frequency shift
associated to the absorption of a photon with $\lambda=698\,$nm.
The excitation probability in the search-mode scan is only
$\sim30\,\%$ since the actual clock transition is under-sampled
because of the need of covering a large scanning span in reduced
time. The excitation pulse duration in this high-resolution mode
is chosen to realize an effective $\pi$ pulse for each
configuration of magnetic field and probe intensity. For this
purpose the actual atomic Rabi frequency is measured through the
observation of the Rabi oscillations (see Fig.
\ref{fig:Rabi_oscillations}). On the carrier the excitation
probability approaches the $70$ \% level. We find that the fitted
Rabi frequency is about $50$ \% of the one calculated from Eq.
\ref{eq:Rabi}, an effect we attribute, together with the reduced
excitation probability of $70$ \%, to the inhomogeneous
distribution of the Rabi frequency among the atoms that could be
given by a residual spatial inhomogeneity of the clock probe, by a
residual misalignment between the probe and lattice axes
(estimated to ~1 mrad) and by the thermal transverse atomic motion
in the lattice potential \cite{Blatt2006}. This effect can be
significantly reduced by employing a probe beam with larger
diameter and a colder atomic sample.
By reducing both the mixing magnetic field and the clock probe
intensity the low drift rate of the clock laser (few
$\text{mHz}/\text{s}$) allowed us to reliably measure Rabi
oscillations lasting for more than $100\,$ms and Fourier limited
spectra below the $10\,$Hz FWHM linewidth (see Fig.
\ref{fig:Rabi_oscillations}).
%These results are achieved by employing a commercial tapered
%amplified diode laser, unstabilized in power and frequency, thus
%providing a demonstration that this technology does not limit the
%atomic coherent evolution on a time scale of more than 100$\,$ms
%and is suitable for lattice clock spectroscopy at the level of
%$\sim10\,$Hz linewidth.
\begin{figure}
\includegraphics[width=0.49\textwidth]{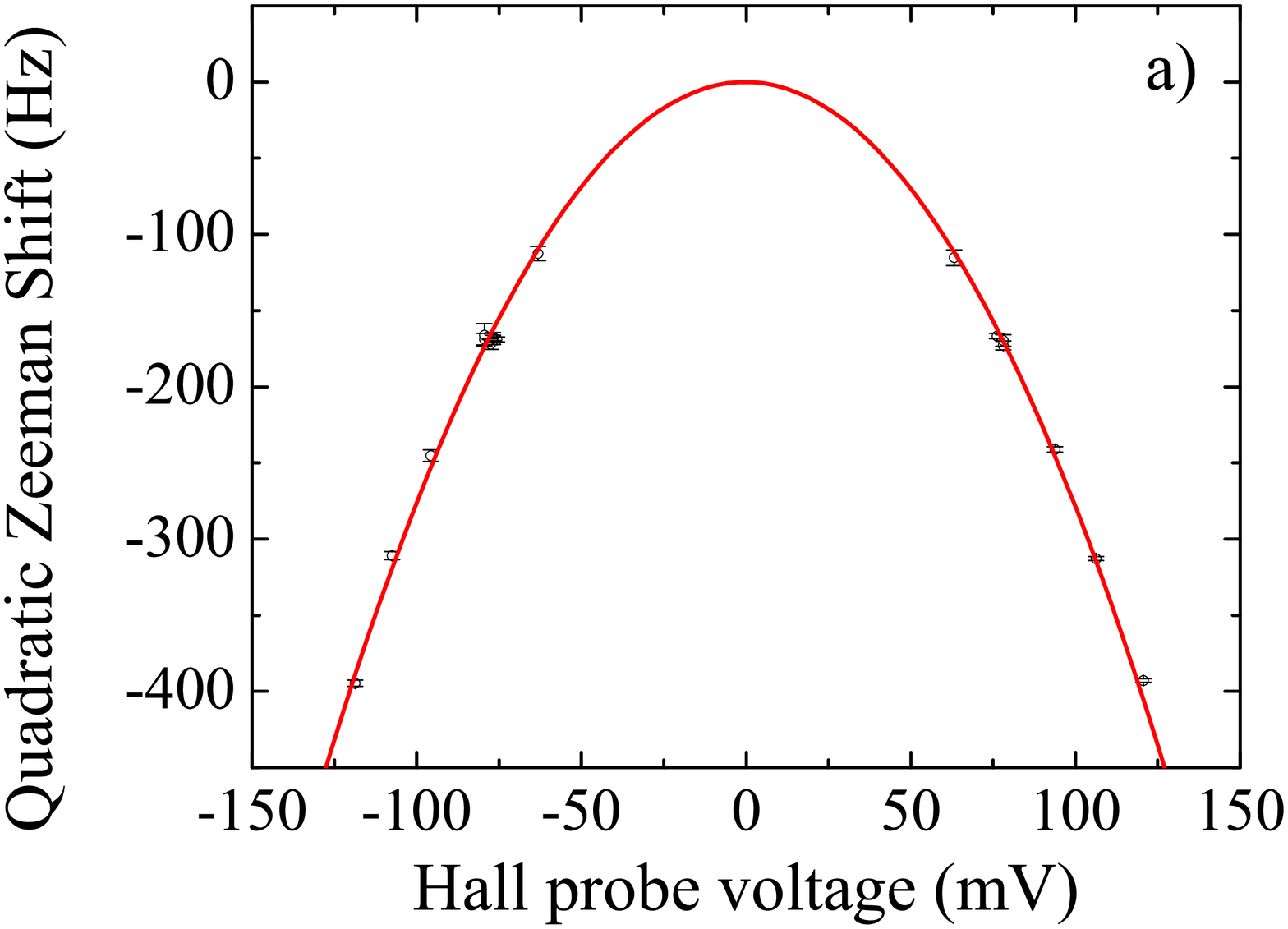}
\includegraphics[width=0.4\textwidth]{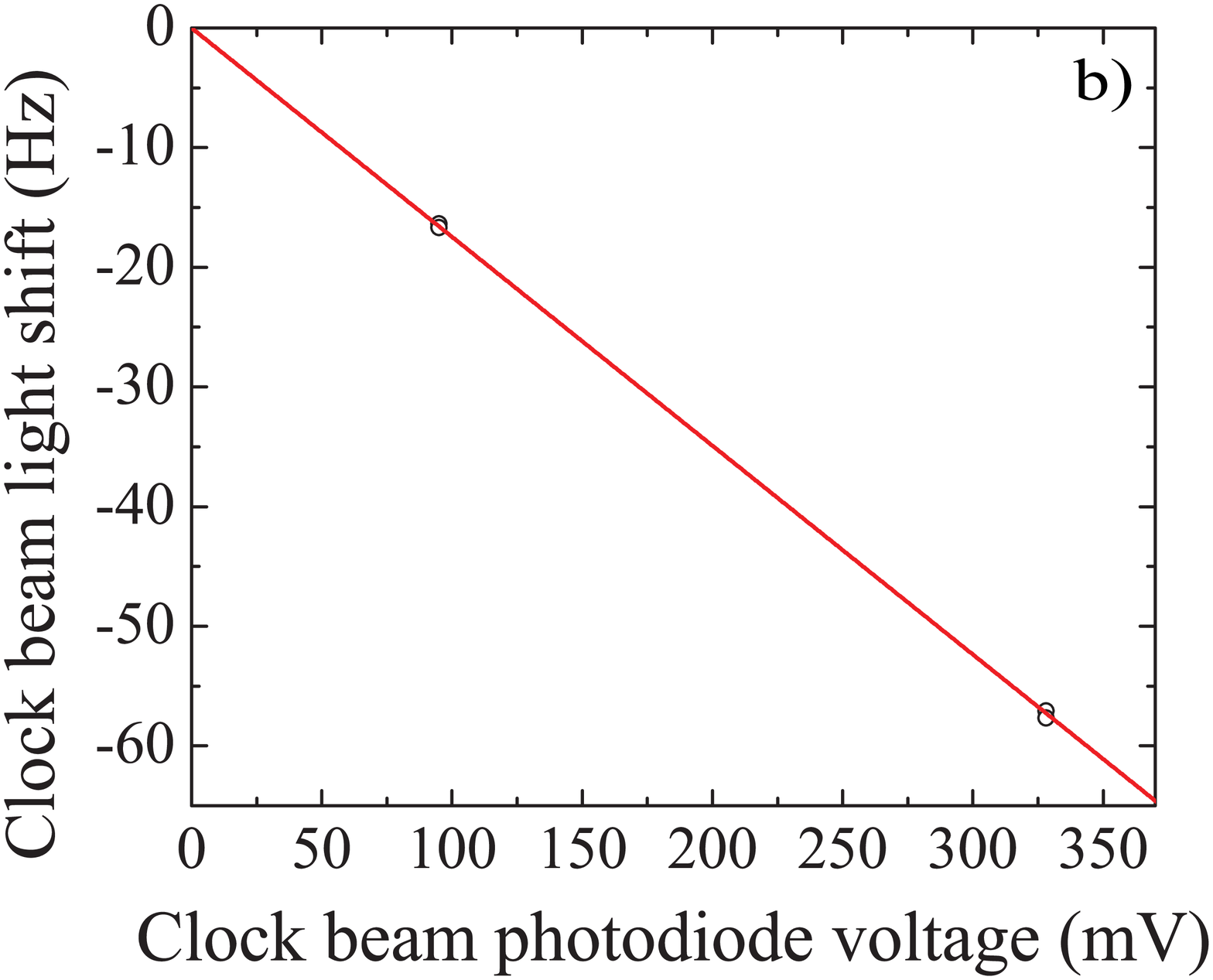}
\vspace{0.1cm}
\includegraphics[width=0.49\textwidth]{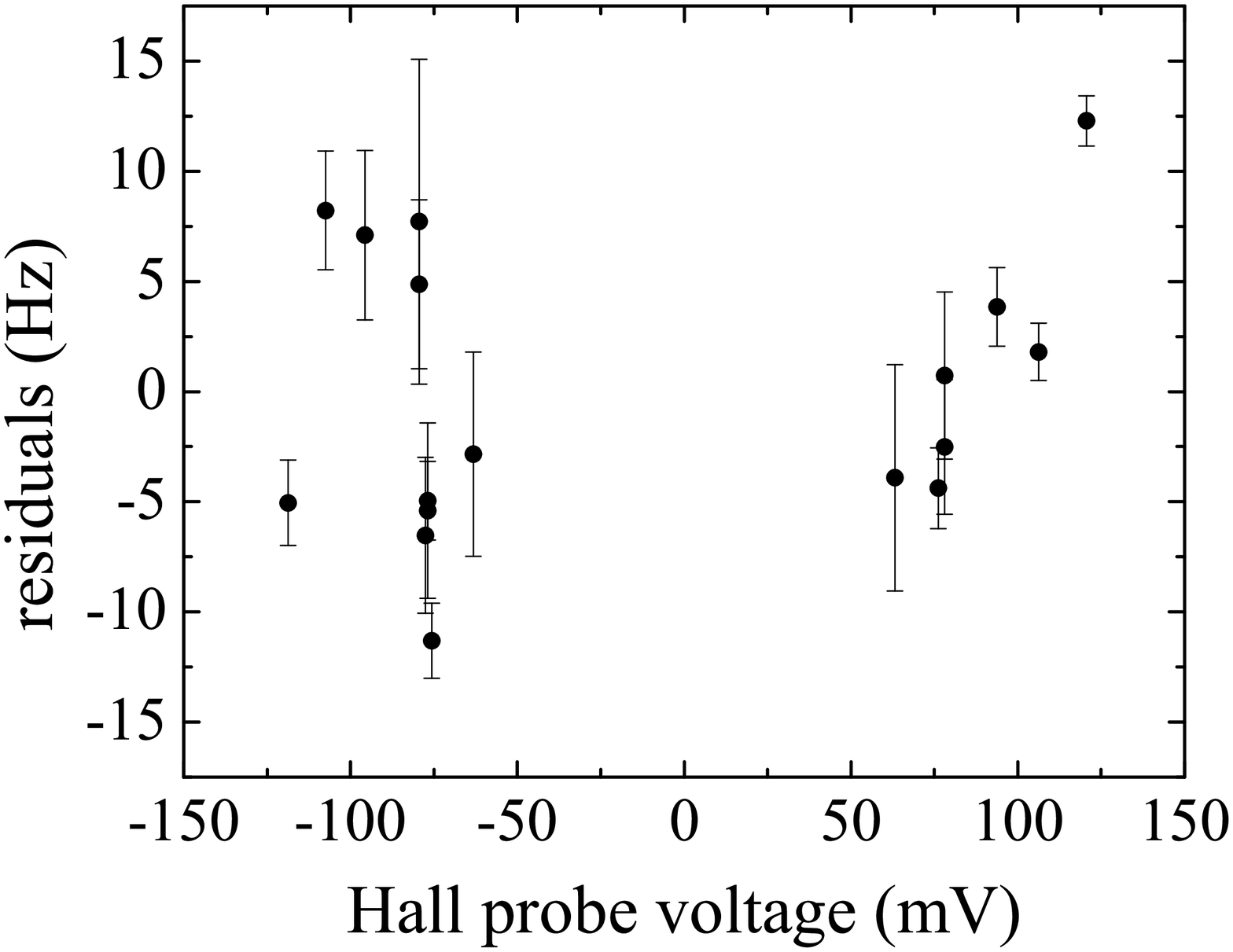}
\includegraphics[width=0.45\textwidth]{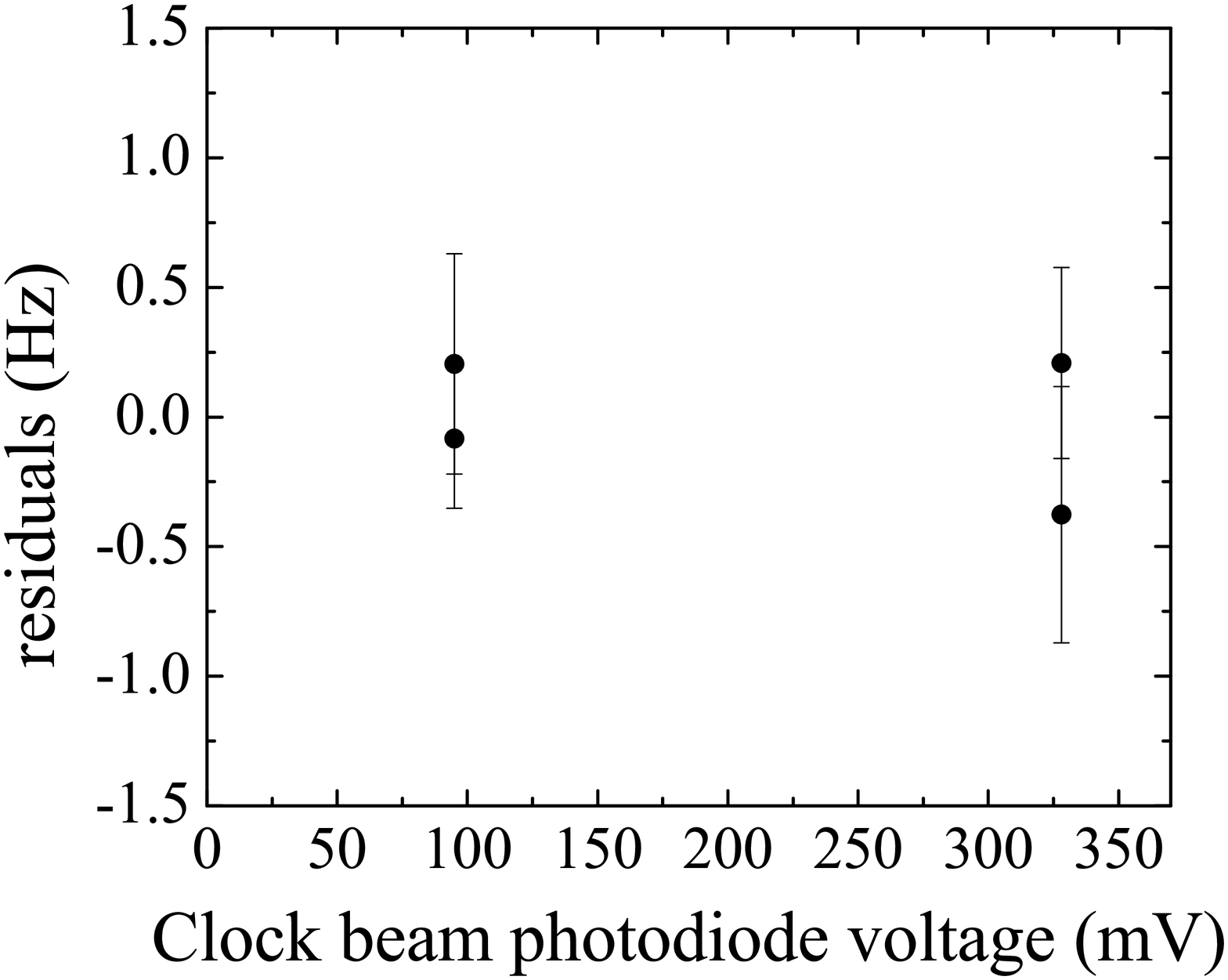}
\caption{\label{fig:shifts} Systematic shifts of the
$^1$S$_0\,$-${}^3$P$_0$ clock transition in $^{88}$Sr optical
lattice clock. a) Measurement of the second order Zeeman shift as
a function of the Hall probe voltage.
%employed for monitoring the applied magnetic field.
%b) Calibration of the Hall probe through
%the conversion of the experimental Zeeman shift into external
%magnetic field by means of the known coefficient
%$\beta=-23.3\,\text{Hz}/\text{mT}^2$.
b) Measurement of the probe AC linear Stark shift as a function of
the photodiode voltage.
%used for controlling the optical power ofthe clock beam
 The respective curves are used to calibrate the Hall probe and the
photodiode to estimate the applied magnetic field $\mathbf{B}$ and
the probe light intensity $I$ through the knowledge of the shift
coefficients \cite{Taichenachev2006}. The two calibration
coefficients are $\gamma=29.0(4)$ mV/mT for the Hall probe and
$\eta=$ 104(1) mV/(W cm$^{-2}$) for the photodiode, respectively.}
%d) Calibration of the photodiode.} .
%through the
%conversion of the experimental ac Stark shift into probe light
%intensity by means of the known coefficient
%$k=-18\,\text{Hz}/(\text{W}/\text{cm}^2)$.}
\end{figure}

A study of the systematics has been carried out and values of
estimated shifts on the $^1$S$_0\,$-${}^3$P$_0$ clock transition
are reported in Table \ref{tab:final_budget}.

The 2$^{\mathrm{nd}}$ order Zeeeman and probe Stark shifts have
been evaluated by scanning around the clock transition and
interleaving measurements with different values of the magnetic
field and probe power. As frequency reference we rely on the clock
cavity resonance whose typical 1 Hz/s linear drift was compensated
to better than 10 mHz/s with a programmable feed-forward AOM
driver.

Magnetic field and clock probe power are monitored at $1\,\%$
precision level with a Hall probe and a photodiode, respectively.
They are calibrated through the known coefficient $\beta$ and $k$
\cite{Taichenachev2006} by measuring the clock frequency shift as
a function of the magnetic field ($2.2-4.1\,$mT range, also
inverting the direction of the field) and probe intensity
($0.9-3.2\,\text{W}/\text{cm}^2$ range), as shown in Fig.
\ref{fig:shifts}. Values are calculated for the $8.0(0.1)\,$Hz
linewidth clock transition (see Table \ref{tab:final_budget}),
obtained with a magnetic field of
$\left|\mathbf{B}\right|=1.19(4)\,$mT and probe beam intensity of
$I=0.922(13)\,\text{W}/\text{cm}^2$, corresponding to a second
order Zeeman shift and probe AC Stark shift of
$\Delta_B=\beta\left|\mathbf{B}\right|^2=-33.5(2.4)\,$Hz and
$\Delta_L=kI=-16.6(2)\,$Hz respectively, where
$\beta=-23.3\,\text{Hz}/\text{mT}^2$ and
$k=-18\,\text{Hz}/(\text{W}/\text{cm}^2)$ \cite{Taichenachev2006}.
The uncertainties on the second order Zeeman shift and probe light
shift are given by the quadratic sum of the standard error from
the fit and the error associated by the voltage reading on the
Hall probe and the photodetector.

%The related uncertainties are extrapolated by measuring the shifts as a function of magnetic field ($3-13\,$mT interval) and clock beam %intensity ($0.9-3.2\,\text{W}/\text{cm}^2$ interval), controlled respectively with an Hall probe and a photodiode with a relative precision %error of $1\%$.
\begin{table}[t]
\caption{\label{tab:final_budget} Systematic frequency shift and
uncertainty budget for the $^{88}$Sr optical lattice clock as
extracted from spectra of the clock transition. The values are
reported for the operating conditions of the $8.0(0.1)\,$Hz clock
transition linewidth (Fig. \ref{fig:clock_spectra} b)).}
\begin{ruledtabular}
\begin{tabular}{lcc}
Contributor & Shift (Hz) & Uncertainty (Hz)\\
\hline
2nd order Zeeman & -33.5 & 2.4\\
Clock light & -16.6 & 0.2\\
Collisions & 1.0 & 0.4\\
Blackbody radiation & -2.5 & 0.5\\
AC Stark (lattice) & -0.7  & 0.9\\
\hline
Total Uncertainty & & 2.8\\
\end{tabular}
\end{ruledtabular}
\end{table}

The effects of collisions on the $^{88}$Sr clock transition have
been studied in detail in \cite{Lisdat2008}. In our optimal
conditions the number of atoms trapped into the lattice is kept at
$7.0(7)\times10^3$ ($\sim7$ atoms per site), so that the
corresponding peak atomic density per lattice site of
$\rho=1.39(14)\times10^{10}\,\text{cm}^{-3}$ leads to a
collisional shift and broadening of $\rho\times
\Delta\nu_{\rho}=1.0(4)\,$Hz and $\rho\times
\gamma_{\text{dep}}/\pi=1.4(6)\,$Hz respectively, with the
coefficients
$\Delta\nu_{\rho}=(7.2\pm2.0)\times10^{-11}\,\text{Hz}/\text{cm}^3$
and $\gamma_{\text{dep}}=(3.2\pm1.0)\times10^{-10}
\text{cm}^3/\text{s}$ \cite{Lisdat2008}. The uncertainties on
collisional shift and broadening take into account our
experimental $10\,\%$ fluctuation of the number of atoms
shot-to-shot and the uncertainties of the $\Delta\nu_{\rho}$ and
$\gamma_{\text{dep}}$ coefficients and density determination.

During the clock spectroscopy measurement the vacuum chamber
temperature is monitored with one thermistor directly attached to
the main chamber. No water cooling is employed to cool the MOT
coils and after a warm up of about three hours, the temperature of
the whole cell stabilizes to $318\,$K. While a more detailed study
of temperature gradients is mandatory for accuracy level below the
10$^{-16}$ level, this lies beyond the scope of the present
article. We estimated the blackbody radiation shift to $-2.5\,$Hz
with an uncertainty of 0.5 Hz, including the uncertainty due to
temperature gradients up to $\pm$ 10 K inside the main cell. The
AC Stark shift induced by the lattice light has been evaluated
considering the effect of the detuning of 0.40 pm with respect to
the magic wavelength for $^{88}$Sr (with a scalar coefficient $8$
Hz/nm/$E_R$). Taking also into account the uncertainty in the
knowledge of the magic wavelength and the smaller contribution
from hyperpolarizability effect \cite{Brusch2006}, the shift
amount to 0.3(4) Hz. The amplified spontaneous emission (ASE) from
the tapered amplifier used for lattice laser is about 40 nm wide
and symmetric around the emission wavelength. With a typical ASE
intensity 40 dB below the carrier, we estimated an additional
shift of -1.0(5) Hz due to this effect. The total value for the AC
Stark shift has been evaluated to be -0.7(9)Hz. At $8\,$Hz clock
transition linewidth the achieved total fractional uncertainty is
$\sim7\times10^{-15}$, with a transition quality factor of
$Q\sim5\times10^{13}$. With a current clock cycle time of $T_c=1$
s (mainly limited by technical delays in the absorption imaging
detection) and an interrogation time $\Delta t=130$ ms, the
stability of the clock is limited at $4\times 10^{-15}$ at 1 s by
the Dick effect \cite{Quessada2003}. With minor changes to the
detection system a reduction of the clock cycle time to $\sim 400$
ms can be implemented, maintaining the same number of atoms into
the lattice, thus reducing the contribution of the Dick noise
below $2\times 10^{-15}$ level.

\section{\label{sec:conclusions} Conclusions}
We presented a compact and transportable optical clock based on
strontium atoms. Novel design solutions allowed us also to reduce
the volume, weight and power consumption with respect to
traditional laser cooling apparatus. As a result the experimental
physics package is contained in a volume of less than 2 m$^3$ and
no water cooling is needed to operate the clock. Furthermore, a
modular architecture ensured a high degree of operation
reliability of the apparatus both in stationary condition and
after a transportation of the experimental set up. To ensure high
clock frequency stability, cooling and trapping stages have been
optimized to allow high efficiency transfer among different
cooling and trapping stages, thus allowing for faster clock cycle
time with high duty cycle. Spectroscopy on $^1$S$_0$ -$^3$P$_0$
clock transition on bosonic $^{88}$Sr isotopes has been
demonstrated with an 8 Hz resolution. Eventually, an evaluation of
the main systematic frequency shifts on the clock transition has
been done and the fractional uncertainty of the clock is
$7\times10^{-15}$.

%\begin{acknowledgments}

The authors acknowledge financial support from ESA and the
European Union Seventh Framework Programme (FP7/2007-2013 grant
agreement 263500, project ``Space Optical Clocks'') and the
European Metrology Research Programme (EMRP) under IND14. The EMRP
is jointly funded by the EMRP participating countries within
EURAMET and the European Union. We also acknowledge support by the
DFG RTG 1729 ``Fundamentals and applications of ultra-cold
matter''. We thank D. V. Sutyrin for useful discussions.
%\end{acknowledgments}
%\bibliographystyle{unsrt}
%\bibliography{TransportableSr}

\begin{thebibliography}{10}

\bibitem{Poli2013}
N.~Poli, C.~W. Oates, P.~Gill, and G.~M. Tino.
\newblock Optical atomic clocks.
\newblock {\em Rivista del Nuovo Cimento}, 36:555--624, 2013.

\bibitem{Bloom2014}
B.~J. Bloom, T.~L. Nicholson, J.~R. Williams, S.~L. Campbell, M.~Bishof,
  X.~Zhang, W.~Zhang, S.~L. Bromley, and J.~Ye.
\newblock An optical lattice clock with accuracy and stability at the
  10$^{-18}$ level.
\newblock {\em Nature}, 506:71--75, 2014.

\bibitem{Chou2010}
C.~W. Chou, D.~B. Hume, J.~C.~J. Koelemeij, D.~J. Wineland, and T.~Rosenband.
\newblock Frequency comparison of two high-accuracy {Al$^+$} optical clocks.
\newblock {\em Phys. Rev. Lett.}, 104(7):070802, 2010.

\bibitem{Hinkley2013}
N.~Hinkley, J.~A. Sherman, N.~B. Phillips, M.~Schioppo, N.~D. Lemke, K.~Beloy,
  M.~Pizzocaro, C.~W. Oates, and A.~D. Ludlow.
\newblock An atomic clock with $10^{-18}$ instability.
\newblock {\em Science}, 341(6151):1215--1218, 2013.

\bibitem{Schiller2009}
S.~Schiller, G.~M. Tino, P.~Gill, C.~Salomon, U.~Sterr, E.~Peik, A.~Nevsky,
  A.~Goerlitz, D.~Svehla, G.~Ferrari, N.~Poli, L.~Lusanna, H.~Klein,
  H.~Margolis, P.~Lemonde, P.~Laurent, G.~Santarelli, A.~Clairon, W.~Ertmer,
  E.~Rasel, J.~Mueller, L.~Iorio, C.~Laemmerzahl, H.~Dittus, E.~Gill,
  M.~Rothacher, F.~Flechner, U.~Schreiber, V.~Flambaum, Wei-Tou Ni, Liang Liu,
  Xuzong Chen, Jingbiao Chen, Kelin Gao, L.~Cacciapuoti, R.~Holzwarth, M.~P.
  Hess, and W.~Schaefer.
\newblock Einstein gravity explorer-a medium-class fundamental physics mission.
\newblock {\em Exp. Astron.}, 23(2):573--610, 2009.

\bibitem{Wolf2009}
P.~Wolf, Ch.~J. Borde, A.~Clairon, L.~Duchayne, A.~Landragin, P.~Lemonde,
  G.~Santarelli, W.~Ertmer, E.~Rasel, F.~S. Cataliotti, M.~Inguscio, G.~M.
  Tino, P.~Gill, H.~Klein, S.~Reynaud, C.~Salomon, E.~Peik, O.~Bertolami,
  P.~Gill, J.~Paramos, C.~Jentsch, U.~Johann, A.~Rathke, P.~Bouyer,
  L.~Cacciapuoti, D.~Izzo, P.~De~Natale, B.~Christophe, P.~Touboul, S.~G.
  Turyshev, J.~Anderson, M.~E. Tobar, F.~Schmidt-Kaler, J.~Vigue, A.~A. Madej,
  L.~Marmet, M.~C. Angonin, P.~Delva, P.~Tourrenc, G.~Metris, H.~Mueller,
  R.~Walsworth, Z.~H. Lu, L.~J. Wang, K.~Bongs, A.~Toncelli, M.~Tonelli,
  H.~Dittus, C.~Laemmerzahl, G.~Galzerano, P.~Laporta, J.~Laskar, A.~Fienga,
  F.~Roques, and K.~Sengstock.
\newblock Quantum physics exploring gravity in the outer solar system: the
  sagas project.
\newblock {\em Exp. Astron.}, 23(2):651--687, 2009.

\bibitem{Chou2010a}
C.~W. Chou, D.~B. Hume, T.~Rosenband, and D.~J. Wineland.
\newblock Optical clocks and relativity.
\newblock {\em Science}, 329(5999):1630--3, 2010.

\bibitem{Fortier2011}
T.~M. Fortier, M.~S. Kirchner, F.~Quinlan, J.~Taylor, J.~C. Bergquist,
  T.~Rosenband, N.~Lemke, A.~Ludlow, Y.~Jiang, C.~W. Oates, and S.~A. Diddams.
\newblock Generation of ultrastable microwaves via optical frequency division.
\newblock 5(7):425--429, 2011.

\bibitem{Rogers1983}
A.~E. Rogers, R.~J. Cappallo, H.~F. Hinteregger, J.~I. Levine, E.~F. Nesman,
  J.~C. Webber, A.~R. Whitney, T.~A. Clark, C.~Ma, J.~Ryan, B.~E. Corey, C.~C.
  Counselman, T.~A. Herring, I~I Shapiro, C.~A. Knight, D.~B. Shaffer, N.~R.
  Vandenberg, R.~Lacasse, R.~Mauzy, B.~Rayhrer, B.~R. Schupler, and J.~C. Pigg.
\newblock Very-long-baseline radio interferometry: The mark {III} system for
  geodesy, astrometry, and aperture synthesis.
\newblock {\em Science}, 219(4580):51--4, 1983.

\bibitem{Leibrandt2011}
D.~R. Leibrandt, M.~J. Thorpe, J.~C. Bergquist, and T.~Rosenband.
\newblock Field-test of a robust, portable, frequency-stable laser.
\newblock {\em Opt. Expr.}, 19(11):10278--10286, 2011.

\bibitem{Predehl2012}
K.~Predehl, G.~Grosche, S.~M.~F. Raupach, S.~Droste, O.~Terra, J.~Alnis, Th.
  Legero, T.~W. H\"{a}nsch, Th. Udem, R.~Holzwarth, and H.~Schnatz.
\newblock A 920-kilometer optical fiber link for frequency metrology at the
  19th decimal place.
\newblock {\em Science}, 336(6080):441--4, 2012.

\bibitem{Williams2008}
P.~A. Williams, W.~C. Swann, and N.~R. Newbury.
\newblock High-stability transfer of an optical frequency over long fiber-optic
  links.
\newblock {\em J. Opt. Soc. Am. B-Opt. Phys.}, 25(8):1284--1293, 2008.

\bibitem{Calonico2014}
D.~Calonico, E.~K. Bertacco, C.~E. Calosso, C.~Clivati, G.~A. Costanzo,
  A.~Godone, M.~Frittelli, A.~Mura, N.~Poli, D.~V. Sutyrin, G.~M. Tino,
  M.~Zucco, and F.~Levi.
  \newblock High accuracy coherent optical frequency transfer over a doubled 642 km fiber link
\newblock {\em arXiv:1404.0395}, in publication on \newblock {\em Appl. Phys. B} DOI: 10.1007/s00340-014-5917-8, 2014.

\bibitem{Bize2005}
S.~Bize, P.~Laurent, M.~Abgrall, H.~Marion, I.~Maksimovic, L.~Cacciapuoti,
  J.~Grunert, C.~Vian, F.P. dos Santos, P.~Rosenbusch, P.~Lemonde,
  G.~Santarelli, P.~Wolf, A.~Clairon, A.~Luiten, M.~Tobar, and C.~Salomon.
\newblock Cold atom clocks and applications.
\newblock {\em J. Phys. B-At. Mol. Opt. Phys.}, 38(9, SI):S449--S468, 2005.

\bibitem{Poli2007}
N.~Poli, R.~E. Drullinger, M.~G. Tarallo, G.~M. Tino, and
M.~Prevedelli.
\newblock Strontium optical lattice clock with all semiconductor sources.
\newblock In {\em Proceedings of the 2007 IEEE International Frequency Control
  Symposium-jointly with the 21st European Frequency and Time Forum, Vols 1-4},
  pages 655--658, 2007.

\bibitem{Poli2009}
N.~Poli, M.~G. Tarallo, M.~Schioppo, C.~W. Oates, and G.~M. Tino.
\newblock A simplified optical lattice clock.
\newblock {\em Applied Physics B}, 97:27, 2009.

\bibitem{Katori2009}
H.~Katori, K.~Hashiguchi, E.~Yu. Il'inova, and V.~D. Ovsiannikov.
\newblock Magic wavelength to make optical lattice clocks insensitive to atomic
  motion.
\newblock {\em Phys. Rev. Lett.}, 103:153004, 2009.

\bibitem{SchioppoPhDThesis}
M.~Schioppo.
\newblock {\em Development of a Transportable Strontium Optical Clock}.
\newblock PhD thesis, Dipartimento di Fisica e Astronomia, Universit\`a di
  Firenze, Firenze, December 2010.

\bibitem{Schioppo2012}
M.~Schioppo, N.~Poli, M.~Prevedelli, S.~Falke, Ch. Lisdat, U.~Sterr, and G.~M.
  Tino.
\newblock A compact and efficient strontium oven for laser-cooling experiments.
\newblock {\em Rev. Sci. Instrum.}, 83(10, 1), 2012.

\bibitem{Baillard2006}
X.~Baillard, A.~Gauguet, S.~Bize, P.~Lemonde, Ph. Laurent, A.~Clairon, and
  P.~Rosenbusch.
\newblock Interference-filter-stabilized external-cavity diode lasers.
\newblock {\em Opt. Commun.}, 266(2):609--613, 2006.

\bibitem{Falke2012}
S.~Falke, M.~Misera, U.~Sterr, C.~Lisdat
\newblock Delivering pulsed and phase stable light to atoms of an optical clock
\newblock {\em Applied Physics B}, 107(2):301--311, 2012.

\bibitem{Vogt2011}
S.~Vogt, C.~Lisdat, T.~Legero, U.~Sterr, I.~Ernsting, A.~Nevsky, and
  S.~Schiller.
\newblock Demonstration of a transportable 1 {Hz}-linewidth laser.
\newblock {\em Appl. Phys. B-Lasers Opt.}, 104(4):741--745, 2011.

\bibitem{Xu2003}
X.~Xu, T.~H. Loftus, J.~L. Hall, A.~Gallagher, and J.~Ye.
\newblock Cooling and trapping of atomic strontium.
\newblock {\em J. Opt. Soc. Am. B}, 20(5):968--976, 2003.

\bibitem{Dinneen1999}
T. P.~Dinneen, K. R.~Vogel, E.~Arimondo, J. L.~Hall, and
A.~Gallagher
\newblock Cold collisions of Sr*-Sr in a magneto-optical trap.
\newblock {\em Phys. Rev. A}, 59:1216, 1999.

\bibitem{Katori1999}
H.~Katori, T.~Ido, Y.~Isoya, and M.~Kuwata-Gonokami.
\newblock Magneto-optical trapping and cooling of strontium atoms down to the photon recoil temperature.
\newblock {\em Phys. Rev. Lett.}, 82:1116, 1999.

\bibitem{Loftus2004}
T.~H. Loftus, T.~Ido, M.~M. Boyd, A.~D. Ludlow, and J.~Ye.
\newblock {Narrow line cooling and momentum-space crystals}.
\newblock {\em Physical Review A}, 70:063413, 2004.

\bibitem{Akatsuka2010}
T.~Akatsuka, M.~Takamoto, and H.~Katori.
\newblock Three-dimensional optical lattice clock with bosonic {Sr}88 atoms.
\newblock {\em Phys. Rev. A}, 81:023402, 2010.

\bibitem{Lisdat2008}
Ch. Lisdat, J.~S. R.~Vellore Winfred, T.~Middelmann, F.~Riehle, and U.~Sterr.
\newblock Collisional losses, decoherence, and frequency shifts in optical
  lattice clocks with bosons.
\newblock {\em Phys. Rev. Lett.}, 103(9), 2009.

\bibitem{Tarallo2011}
M.~G. Tarallo, N.~Poli, M.~Schioppo, D.~Sutyrin, and G.~M. Tino.
\newblock A high-stability semiconductor laser system for a {Sr}-88-based
  optical lattice clock.
\newblock {\em Appl. Phys. B-Lasers Opt.}, 103(1):17--25, 2011.

\bibitem{Taichenachev2006}
A.~V. Taichenachev, V.~I. Yudin, C.~W. Oates, C.~W. Hoyt, Z.~W. Barber, and
  L.~Hollberg.
\newblock Magnetic field-induced spectroscoy of forbidden optical transitions
  with application to lattice-based optical atomic clocks.
\newblock {\em Phys. Rev. Lett.}, 96:083001, 2006.

\bibitem{Blatt2006}
S.~Blatt, J.~W. Thomsen, G.~K. Campbell, A.~D. Ludlow, M.~D. Swallows, M.~J.
  Martin, M.~M. Boyd, and J.~Ye.
\newblock Rabi spectroscopy and excitation inhomogeneity in a one-dimensional
  optical lattice clock.
\newblock {\em Phys. Rev. A}, 80:052703, 2009.

\bibitem{Brusch2006}
A.~Brusch, R.~Le~Targat, X.~Baillard, M.~Fouch\'{e}, and P.~Lemonde.
\newblock Hyperpolarizability effects in a {Sr} optical lattice clock.
\newblock {\em Phys. Rev. Lett.}, 96:103003, 2006.

\bibitem{Quessada2003}
A.~Quessada, R. P.~Kovacich, I.~Courtillot, A.~Clairon,
G.~Santarelli, and P.~Lemonde
\newblock The Dick effect for an optical frequency standard
\newblock {\em J. Opt. B: Quantum Semiclass. Opt.}, 5:S150, 2003.

\end{thebibliography}

\end{document}